\documentclass[10pt,conference,letterpaper]{IEEEtran}
\IEEEoverridecommandlockouts
\usepackage{cite}
\usepackage{amsmath,amssymb,amsfonts}
\usepackage{algorithmic}
\usepackage{graphicx}
\usepackage{textcomp}
\usepackage{xcolor}
\def\BibTeX{{\rm B\kern-.05em{\sc i\kern-.025em b}\kern-.08em
    T\kern-.1667em\lower.7ex\hbox{E}\kern-.125emX}}

\usepackage{soul}

\usepackage{bbm}
\usepackage{graphicx}

\usepackage{amsmath,amssymb,amsfonts}
\usepackage{algorithmic}
\usepackage{graphicx}
\usepackage{textcomp}
\usepackage{xcolor}
\usepackage[ruled,vlined,linesnumbered]{algorithm2e}

\usepackage{mathrsfs}
\usepackage{graphics}
\usepackage{epsfig}
\usepackage{mathptmx}
\usepackage{times}
\usepackage{amssymb}
\usepackage{graphicx}
\usepackage{color}
\usepackage{float}
\usepackage{amsfonts}
\usepackage{amsthm}
\usepackage{indentfirst}
\usepackage{tablefootnote}	
\usepackage{enumerate}
\usepackage{hyperref}
\usepackage{array}
\usepackage{booktabs} 

\usepackage{multirow}
\usepackage{cite}
\usepackage{enumerate}
\usepackage{mathrsfs}
\usepackage{graphicx}
\usepackage{subcaption}

\hypersetup{hidelinks,
 colorlinks=true,
 allcolors=black,
 pdfstartview=Fit,
 urlcolor=magenta,
 breaklinks=true}

\definecolor{mygrey}{gray}{0.9} 

\newtheorem{definition}{Definition}

\begin{document}

\title{
Continual Learning with Strategic Selection and Forgetting for Network Intrusion Detection
}

\author{
\IEEEauthorblockN{
Xinchen Zhang\IEEEauthorrefmark{2}\IEEEauthorrefmark{3},
Running Zhao\IEEEauthorrefmark{2},
Zhihan Jiang\IEEEauthorrefmark{2},
Handi Chen\IEEEauthorrefmark{2},
Yulong Ding\IEEEauthorrefmark{3}, \\
Edith C.H. Ngai\IEEEauthorrefmark{2}\IEEEauthorrefmark{1},
Shuang-Hua Yang\IEEEauthorrefmark{3}\IEEEauthorrefmark{4}\IEEEauthorrefmark{1}
}

\IEEEauthorblockA{
\IEEEauthorrefmark{2}The University of Hong Kong
\IEEEauthorrefmark{3}Shenzhen Key Laboratory of Safety and Security for Next Generation \\ of Industrial Internet, Southern University of Science and Technology
\IEEEauthorrefmark{4}University of Reading
}

\thanks{\IEEEauthorrefmark{1}Co-corresponding authors: Edith C.H. Ngai (Email: chngai@eee.hku.hk), Shuang-Hua Yang (Email: shuang-hua.yang@reading.ac.uk)}
}

\maketitle
\begin{abstract}
Intrusion Detection Systems (IDS) are crucial for safeguarding digital infrastructure. 
In dynamic network environments, both threat landscapes and normal operational behaviors are constantly changing, resulting in concept drift. While continuous learning mitigates the adverse effects of concept drift, insufficient attention to drift patterns and excessive preservation of outdated knowledge can still hinder the IDS's adaptability. In this paper, we propose SSF (Strategic Selection and Forgetting), a novel continual learning method for IDS, providing continuous model updates with a constantly refreshed memory buffer. Our approach features a strategic sample selection algorithm to select representative new samples and a strategic forgetting mechanism to drop outdated samples. The proposed strategic sample selection algorithm prioritizes new samples that cause the `drifted' pattern, enabling the model to better understand the evolving landscape. Additionally, we introduce strategic forgetting upon detecting significant drift by discarding outdated samples to free up memory, allowing the incorporation of more recent data. SSF captures evolving patterns effectively and ensures the model is aligned with the change of data patterns, significantly enhancing the IDS's adaptability to concept drift. The state-of-the-art performance of SSF on NSL-KDD and UNSW-NB15 datasets demonstrates its superior adaptability to concept drift for network intrusion detection. 
The code is released at \href{https://github.com/xinchen930/SSF-Strategic-Selection-and-Forgetting}{https://github.com/xinchen930/SSF-Strategic-Selection-and-Forgetting}.

\end{abstract}

\begin{IEEEkeywords}
continual learning, concept drift, network intrusion detection
\end{IEEEkeywords}

\section{Introduction}
Intrusion Detection Systems (IDSs) serve as the primary defenders of digital infrastructures and interconnected systems, such as the Internet of Things, playing a critical role in monitoring network traffic to identify and alert on unauthorized or malicious activities \cite{GARCIATEODORO200918,FERRAG2020102419, zhang2024anomaly,idsiot,dai2024sarad,xu2024skip,radio2text,radi2speech}. They act as early warning systems, providing vital defense against the constant threat of cyber-attacks and ensuring system integrity \cite{li2021comprehensive}. 

In dynamic network environments, both the threat landscapes and normal operational behaviors are constantly changing \cite{aoc-ids}. Attackers always look for new vulnerabilities, leading to new attack types. Additionally, user behaviors may shift, resulting in changing normal operation patterns. These elements drive concept drift \cite{gama2014survey}, a phenomenon characterized by changing statistical distribution in data over time \cite{owad}. 
The effectiveness of IDSs depends on their ability to detect, anticipate, and adapt to the above changes. This adaptability is crucial for maintaining their resilience against the persistent evolution of both cyber attacks and normal patterns.

However, traditional learning-based IDSs \cite{feco,conIDS,9837465} typically function with static models that, once trained, remain unchanged regardless of subsequent shifts in the environment, like system behaviors or attack strategies. 
This static nature reveals that most existing learning-based IDSs cannot cope with the inherent variability of dynamic environments, as they lack mechanisms to adapt to concept drift. 

By incorporating continual learning, IDSs can adapt to concept drift with the capability for continuous updates and evolution \cite{pendlebury2019tesseract,jan2020throwing}. To keep pace with concept drifts, it is essential to select new samples that accurately reflect these evolving patterns. 
However, existing continual learning strategies typically employ random sampling \cite{er1,er2} or rely on criteria such as the importance of a sample in the learning process or its representativeness of the overall distribution \cite{owad}. 
These criteria often fail to provide adequate attention to the `drifted' pattern, the most challenging aspect. 
Simply treating the `drift' as part of the overall distribution overlooks its critical role in causing concept drift. 
In addition, the most common solution in current continual learning methods \cite{lwf,ewc,agem,gem} is to preserve as much old knowledge as possible to prevent catastrophic forgetting \cite{french1999catastrophic}, where a model forgets previously learned information upon learning new data. 
However, they ignore the potential concern that past knowledge becomes out-of-date or requires correction when drifting occurs. Maintaining outdated or incorrect knowledge can hinder, rather than facilitate, the model's adaptation to new drifts \cite{ye2022learning}. 

In this work, we present SSF (Strategic Selection and Forgetting), a novel continual learning method that provides dynamic model updates to the IDS, facilitated by a refreshed memory buffer to strategically address concept drifts by including the most representative new data while retaining useful historical data.
More specifically, we propose a novel sample selection strategy that identifies the most representative new samples for the `drifted' pattern. By accurately capturing these drifts, our algorithm significantly enhances the model's adaptability to concept drift and provides valuable insights into the evolving landscape. For the unchanged pattern, we utilize already labeled old samples to save labeling effort. This combination of labeled old samples for the unchanged pattern and unlabeled new samples for the drifted pattern offers a comprehensive and resource-efficient foundation for updating the model. 
Moreover, upon detecting significant drift, we employ strategic forgetting, which discards old samples that no longer represent the new pattern. This timely removal of outdated or potentially incorrect knowledge ensures the model adapts effectively to new patterns \cite{jiang2021crowdpatrol} and frees up space of the limited memory buffer \cite{9459954} for incorporating more recent data. During concept drift, retaining new samples that naturally represent the new pattern can be more beneficial for adaptation than keeping old samples. 

Our contributions are summarized as follows: 
\begin{itemize} 

\item We present SSF, a novel continual learning method for IDSs, featuring a strategic sample selection algorithm and a strategic forgetting mechanism. Our method facilitates effective and accurate adaptation of IDSs in dynamic environments with concept drift due to constantly changing system behaviors and evolving attack strategies.

\item We design a strategic sample selection algorithm that identifies the most representative new samples for the `drifted' pattern.
Our approach excels in capturing evolving trends by prioritizing the underlying causes of concept drift, ensuring the model's high adaptability.

\item We introduce strategic forgetting to actively discard outdated or potentially incorrect knowledge that no longer represents current behaviors, especially when significant concept drift occurs. This facilitates effective adaptation to new patterns by keeping the memory buffer up-to-date.

\item We validate the superior performance and adaptability of our method on network traffic datasets, NSL-KDD and UNSW-NB15. Comparative experiments show that our method consistently outperforms baselines under varying labeling resources. The ablation study highlights the individual contributions of each component. 

\end{itemize}

\begin{figure}[!t]
    \centering
    \includegraphics[width=.49\textwidth]{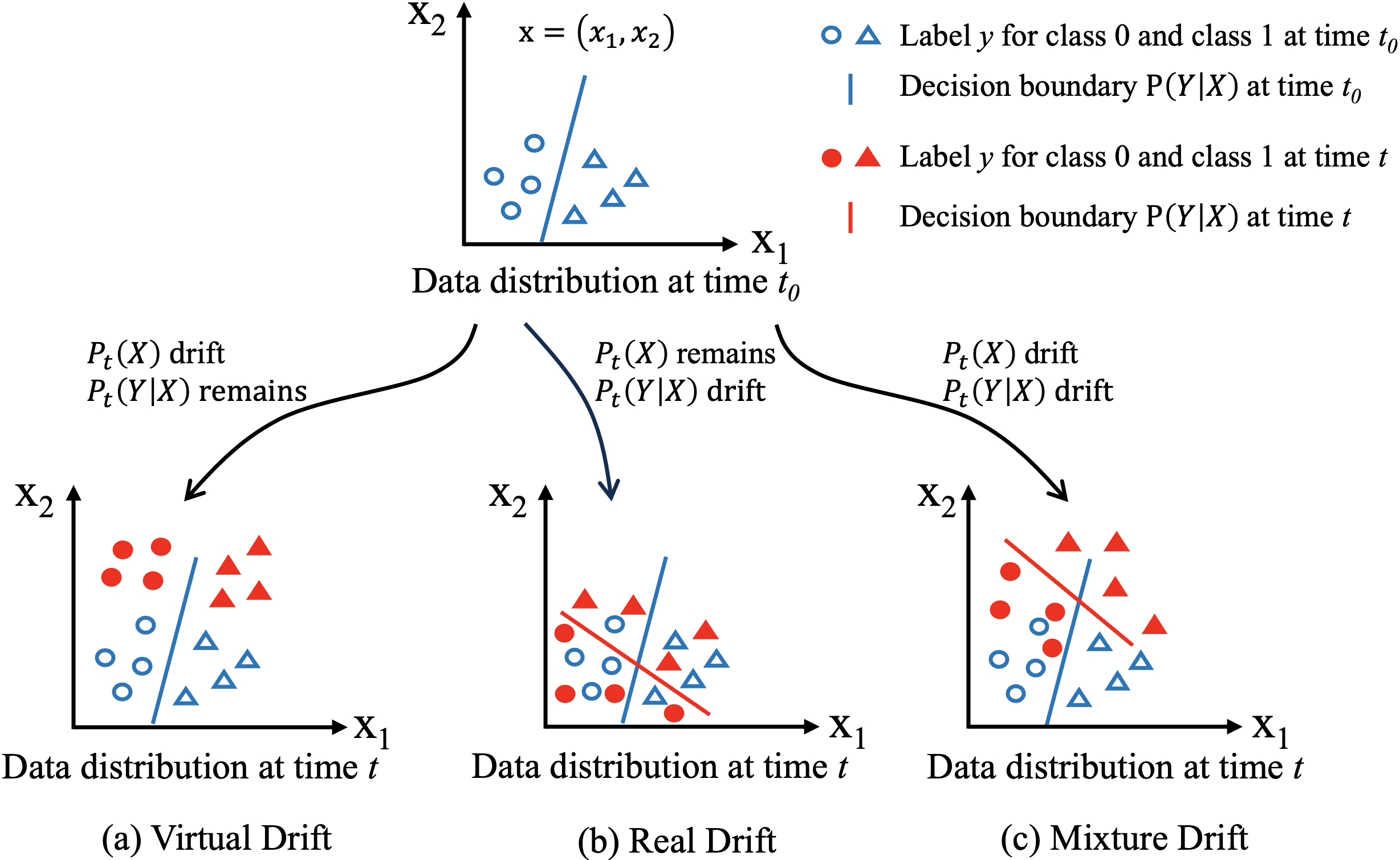}
    \caption{Illustration of concept drift.}
    \label{fig:concept drift}
    \vspace{-0.5cm}
\end{figure}

\section{Preliminaries for Concept Drift}

Learning-based applications often perform well under independently and identically distributed (i.i.d.)  assumption. 
However, in open-world scenarios, this assumption may fail as test distributions can diverge \cite{owad}, causing concept drift. 

\begin{definition}[Concept Drift]
\label{def:info}
    Concept drift refers to the phenomenon where the statistical properties of the joint probability distribution $P(X, Y)$ of the input feature space $X$ and the label space $Y$ change over time. Formally: 
    \begin{equation}
        P(X, Y, t) \neq P(X, Y, t_0), \quad t \neq t_0, 
    \end{equation}
    where \( P(X, Y, t) \) represents the joint probability distribution at time \( t \), and \( t \neq t_0 \) indicates \( t \) and \( t_0 \) are different time points.
\end{definition}

Considering that the joint probability distribution $P(X, Y) = P(X) \times P(Y|X)$, concept drift can be attributed to three sources \cite{lu2018learning}: (1) change of the data distribution $P(X)$, known as virtual shift, shown in \figurename{~\ref{fig:concept drift}}a; (2) change of the conditional probability of the label space $P(Y|X)$, know as actual shift, shown in \figurename{~\ref{fig:concept drift}}b; (3) a mixture of the above two changes, shown in \figurename{~\ref{fig:concept drift}}c.

Since virtual shift does not affect model performance, we focus on actual shift. 
Direct observation of the actual shift in 
\( \mathbb{P}(Y|X) \) is impractical, so we use the model's learned distribution \( \mathbb{P}(f(X) \mid X) \) as an approximation. 
Therefore, addressing concept drift translates to addressing the drift in the learned probabilistic distribution \( \mathbb{P}(f(X) \mid X) \). 

In intrusion detection, \( x \in X \) represents a network traffic sample and \( y \in Y \) represents the label, where \( y = \{0, 1\} \) denotes normal (0) or abnormal (1) traffic. \( f(x) = \mathbb{P}(y=0 \mid x) \) denotes the probability that \( x \) is classified as normal by the model.

\begin{definition}[Concept Drift in Intrusion Detection]
For a dataset \( D \), the learned probabilistic distribution is given by:
\[ 
\{f(x_i) \mid x_i \in D\},
\]
which represents a discrete realization of the probability distribution \( \mathbb{P}(f(X) \mid X) \) based on the data points in \( D \). 

Concept drift in intrusion detection is reflected as a drift in the learned discrete probabilistic distribution between the old dataset \( \{f(x_i) \mid x_i \in D_{t_0}\} \) and the new dataset \( \{f(x_i) \mid x_i \in D_t\} \).
\end{definition}

In the following sections, the term `distribution' refers to this discrete learned probabilistic distribution.

\begin{figure*}[!t]
    \centering
    \includegraphics[width=\textwidth]{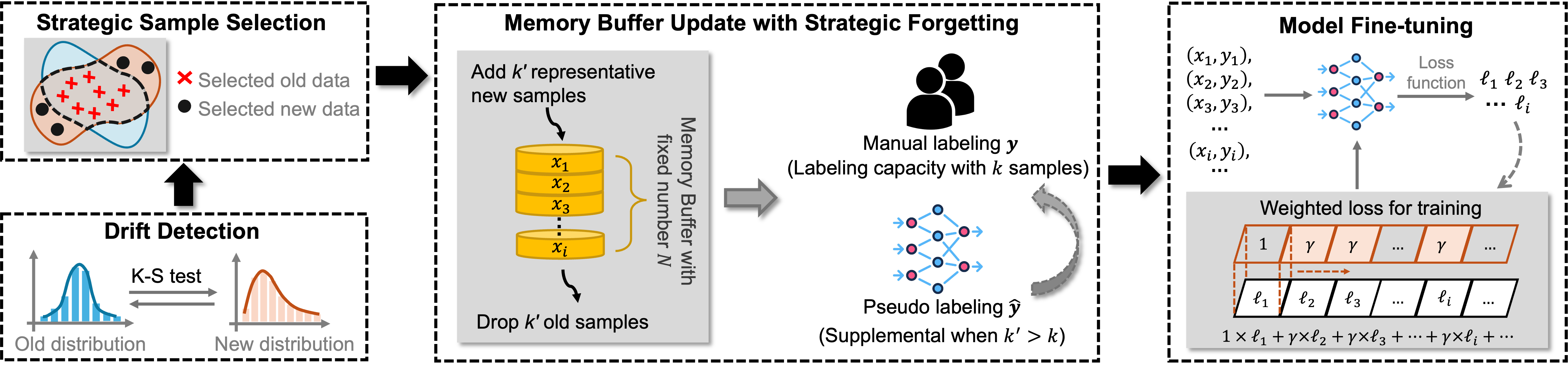}
    \caption{Overview of the proposed continual learning method with strategic selection and forgetting for intrusion detection.}
    \label{fig:overview}
    \vspace{-0.5cm}
\end{figure*}

\section{Methodology}

This section details the framework of SSF, which follows the common logic of memory-based continual learning methods \cite{er1,agem}: training the model on a fixed memory buffer, periodically updating the memory buffer/dataset, and fine-tuning the model based on the updated dataset.
 
As shown in \figurename{~\ref{fig:overview}}, the workflow of SSF comprises four main steps: drift detection, sample selection, memory buffer update, and model fine-tuning, which are executed periodically. The drift detection step identifies significant changes in the distribution.
The sample selection step aims to find the most valuable samples for update training/fine-tuning with minimal labeling resources. 
The memory buffer update step takes different actions based on drift detection results, employing strategic forgetting when drift is detected to help the model adapt quickly to significant changes. Finally, the model fine-tuning step adjusts the model based on the updated memory buffer, ensuring continuous adaptation to new data.

\subsection{Drift Detection}

Considering a scenario where the memory buffer \( M_t = \{(x_i, y_i) \mid i \in \{1, 2, \ldots, N\} \} \) with size \( N \) is updated periodically by new samples/dataset \( D_t = \{(x_i, \emptyset) \mid i \in \{1, 2, \ldots, N'\} \} \), with size \( N' \). For the new dataset \( D_t \), all new samples are initially unlabeled. 
When updating the memory buffer, we can label a subset of new samples under labeling resource constraints.

We initiate the continual learning process with a critical preliminary step: examining whether the new distribution  $P^n = \{f(x_i) \mid x_i \in D_t\}$ of new dataset \( D_t \), has drifted from old distribution $P^o = \{f(x_i) \mid x_i \in M_t\}$ of the existing memory buffer \( M_t \). 
The superscript `o' denotes variables related to old samples, while the superscript `n' for new samples. 
Detecting any such drift is essential, as it sets the stage for subsequent actions based on the drift status.

To detect distributional shifts, we employ the Kolmogorov-Smirnov (K-S) test \cite{berger2014kolmogorov}, which determines if two discrete samples originate from the same continuous distribution, thereby identifying any significant changes in the distribution. In this context, `samples' are subsets of data extracted from a larger population. The K-S test is a nonparametric statistical method, which allows for the detection of distribution drifts without specific assumptions about the underlying distribution. This kind of method provides greater flexibility and robustness, particularly useful in scenarios where the exact distribution is unknown. 
Specifically, the K-S test measures the maximum distance between the empirical cumulative distribution functions of two samples. A large distance suggests the samples may come from different distributions, indicating potential drift between the memory buffer \( M_t \) and the new dataset \( D_t \). 

The K-S test produces a p-value, which measures the likelihood that the observed difference between samples could occur if they were from the same distribution. The drift detection result is determined by comparing the p-value with a predefined significance level, typically set at \( \alpha \). We define a decision function \( h(\cdot) \) for drift detection as follows:
\[
h(p) = 
\begin{cases} 
1 & \text{if } p < \alpha, \text{ drift detected}, \\
0 & \text{if } p \geq \alpha, \text{ no drift detected}.
\end{cases}
\]
A p-value less than the significance level \( \alpha \) indicates a significant distribution drift between the memory buffer \( M_t \) and the new dataset \( D_t \). Conversely, if the p-value is greater than or equal to the threshold, it suggests that there is no significant drift.

\subsection{Strategic Sample Selection}
Before updating the memory buffer by incorporating new samples and discarding old samples, it is essential to identify the most representative samples. Selecting the right samples is critical for effective training, especially when labeling resources and memory resources are limited, as it ensures that the memory buffer contains the most relevant information for the model to learn from with minimal labeling resource cost.

This step has two phases: \textbf{(i) common distribution representation} selects old samples to represent the common distribution between the old and new data. Both old samples and new samples can represent the common distribution, but choosing already labeled old samples helps conserve labeling resources; and \textbf{(ii) distribution drift representation} selects new samples to represent the distribution drift between the old and new data. This ensures that the most representative and meaningful new samples are included, maximizing the value of each labeled sample.

For common distribution representation, both new and old samples can contribute to the construction of the common distribution between the old and new distributions. However, the distribution drift can only be represented by new samples. Therefore, using old samples to reconstruct the new distribution essentially means selecting old samples that represent the common distribution between the old and new distributions.

To achieve this, we introduce a mask vector \( \mathbf{m}^o = \{m^o_1, m^o_2, ..., m^o_i, ...\} \) for the old samples within the memory buffer \( M_t \). Each entry \( m^o_i \) in the mask vector represents the selection decision for the corresponding old sample \( x^o_i \). 
The mask values, ranging between 0 and 1, indicate selecting a sample for representing the common distribution when the value is 0.5 or higher, and not selecting it otherwise. 
We formulate the problem of reconstructing the new distribution with selected old samples $x^o$ as:
\begin{equation}
    \min_{m^o} \mathcal{L}_{\text{old}}, \quad \text{s.t. } m^o_i \in [0, 1],
\end{equation}
\begin{equation}
    \mathcal{L}_{\text{old}} = D_{\text{KL}} \left( \mathcal{H}_\theta (P^n) \big\| \mathcal{H}_\theta (\textbf{m}^o \odot P^o) \right).
\end{equation}
Here, $ \textbf{m}^o \odot P^o$ denotes the distribution constructed by selected old samples with Hadamard (element-wise) product $\odot$. 
$\mathcal{H}_\theta(\cdot)$ converts outputs into $\theta$-bin vectors representing the frequency distribution in a histogram, with $\theta \in \mathbb{N}$ indicating the number of bins.
To measure the `distance' between two distributions, we employ the Kullback-Leibler (KL) divergence $D_{\text{KL}}(P \parallel Q)$, a statistical metric that captures the difference between two distributions. 
The KL divergence is calculated as:
\[ D_{\text{KL}}(P \parallel Q) = \sum_{x \in \mathcal{X}} P(x) \log \frac{P(x)}{Q(x)} . \]

Similarly, for distribution drift representation, we introduce a mask vector $\textbf{m}^n$ for new samples in the new sample dataset \( D_t \). $m^n_i \in [0,1]$ is the indicator of whether or not to select $x^n_i$. 
Thus, we formulate the problem of reconstructing the distribution drift with selected new samples $x^n$ as:
\begin{equation}
    \min_{m^t} \mathcal{L}_{\text{new}}, \quad \text{s.t. } m^n_i \in [0, 1],
\end{equation}
\begin{equation}
    \mathcal{L}_{\text{new}} = D_{KL} \left( \mathcal{H}_\theta \left( P^n \right) \| \mathcal{H}_\theta \left( (\textbf{m}^o \odot P^o) \oplus (\textbf{m}^n \odot P^n) \right) \right).
\end{equation}
Here, \( (\textbf{m}^o \odot P^o) \) is the distribution constructed by selected old samples, and \( (\textbf{m}^n \odot P^n) \) is the distribution constructed by selected new samples. The symbol \(\oplus\) denotes the combination of these two distributions, representing the aggregation of samples from both selected old and new distributions into a single new distribution.

It is important to note that in the process of identifying new samples representing distribution drift, we do not directly use new samples to represent the `drifted' part in the new distribution. This is because the memory buffer and the newly acquired dataset typically differ in size, and simply adding these two distributions of different sizes might not yield the desired overall distribution. Instead, we combine both new and old samples to form the new distribution with fixed old samples to address the issue of differing data volumes between the new and old distributions. 
Since old samples adequately represent the common distribution in the first step of sample selection, the new samples naturally need to represent the `drift' part to minimize $\mathcal{L}_{\text{new}}$ during the second step. 

\begin{figure}[!t]
    \centering
    \includegraphics[width=.35\textwidth]{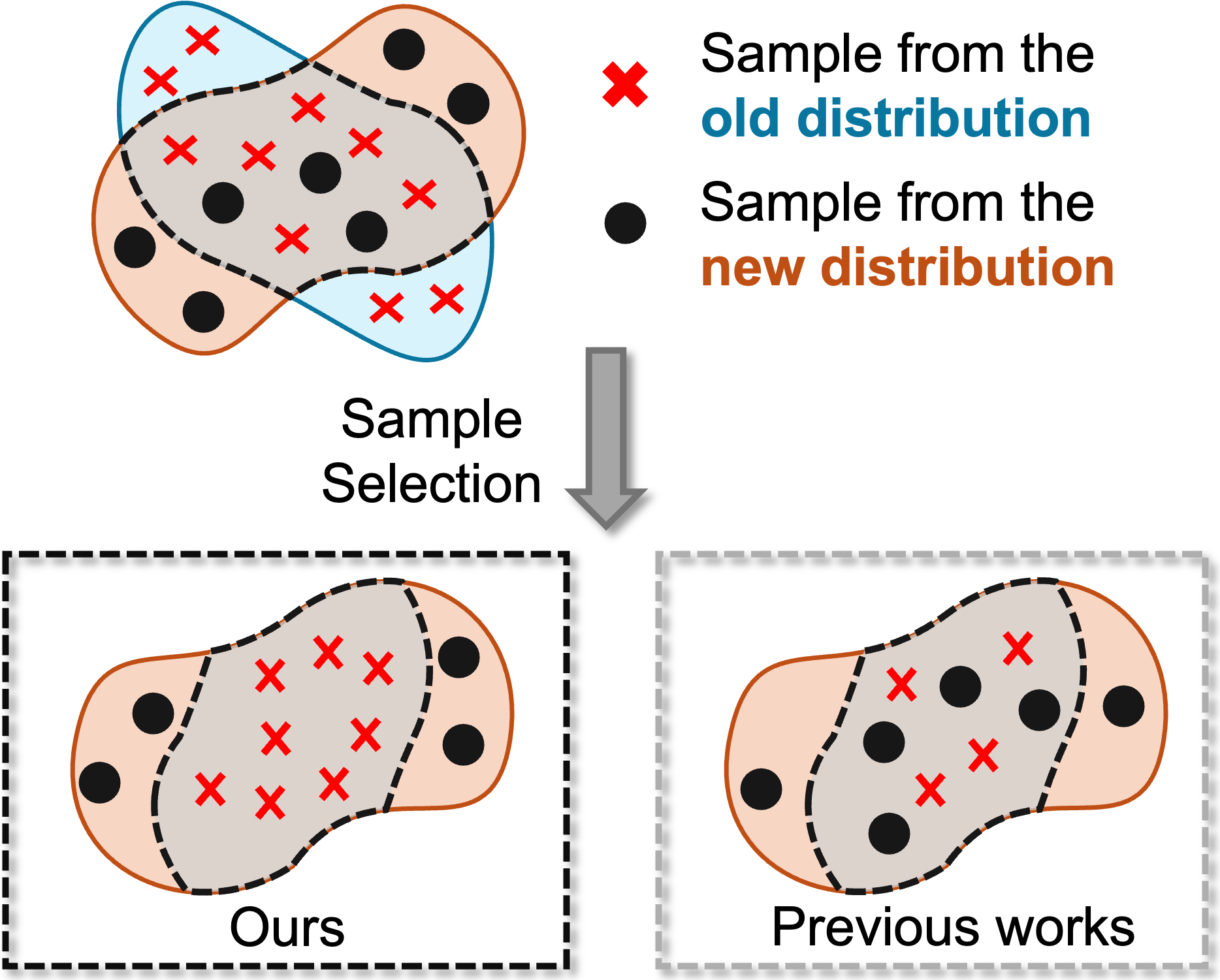}
    \caption{Illustration of our and previous sample selection methods for distribution drift.}
    \label{fig:sample_selection}
    \vspace{-0.5cm}
\end{figure}

\figurename{~\ref{fig:sample_selection}} illustrates the difference between our sample selection method and previous methods. Traditional approaches typically select samples based on the representativeness of the overall distribution. In these methods, both new and old samples from the common region of the old and new distributions are selected simultaneously. Using a new sample that requires labeling to represent a region that can be covered by an old sample (which does not require labeling) is a waste of valuable labeling resources. Furthermore, the drifted part in the overall distribution may be a small but critical region. If samples are selected solely based on the representativeness of the overall distribution, the common distribution with no drift, which occupies the majority of the distribution, is likely to dominate the selection process, leading to insufficient attention on the drifted part crucial for addressing concept drift. 
Our method optimizes labeling resources by having old samples represent the common region and using new samples to represent only the drifted part to capture the drift more effectively.

We employ a gradient descent method to address the above optimization problems. For the initialization, since old samples are already labeled while new samples are not, we aim to retain as many old samples as possible and include as few new samples as necessary to optimize labeling resources. Therefore, we randomly initialize $\mathbf{m}^o$ in the range [0.5, 1] and $\mathbf{m}^n$ in the range [0, 0.5].

\begin{algorithm}[t]
\caption{Memory Buffer Update}
\label{alg:memory_update}
\SetAlgoLined
\SetKwFunction{SelectNewSamples}{SelectNewSamples}
\SetKwFunction{DropOldSamples}{DropOldSamples}
\SetKwFunction{PseudoLabeling}{PseudoLabeling}
\SetKwFunction{UpdateMemoryBuffer}{UpdateMemoryBuffer}
\SetKwFunction{ManualLabeling}{ManualLabeling}
\SetKwFunction{Num}{Num}
\KwIn{Drift detection result $h(p)$, previous memory buffer $M_{t-1}$, new dataset $D_t$, labeling resource $k$, model $f(\cdot)$}
\KwOut{Updated memory buffer $M_t$}

\If{no drift detected: $h(p) = 0$}{
    $X_{t}^{\text{select}} = $ \SelectNewSamples($D_t$, $k$, $m^n$)\; 
    $D_{t}^{\text{select}} = $ \ManualLabeling($X_{t}^{\text{select}}$)\;
    $M_{t-1}^{\text{drop}} = $ \DropOldSamples($M_{t-1}$, $k$, $m^o$)\;
}
\ElseIf{drift detected: $h(p) = 1$}{
    $k' = \text{max}(k, \Num(\mathbf{m}^o < 0.5))$\;
    $M_{t-1}^{\text{drop}} = $ \DropOldSamples($M_{t-1}$, $k'$, $m^o$)\;
    $X_{t}^{\text{select}} = $ \SelectNewSamples($D_t$, $k'$, $m^n$)\;
    \If{\Num($M_{t-1}^{\text{drop}}$) $\leq$ $k$}{
        $D_{t}^{\text{select}} =$ \ManualLabeling($X_{t}^{\text{select}}$)\;
    }
    \Else{
        $D_{t}^{\text{select}} = $ \PseudoLabeling($X_{t}^{\text{select}}$, $f(\cdot)$, $k$)\; 
    }
}

$M_t = (M_{t-1} \setminus M_{t-1}^{\text{drop}}) \cup D_{t}^{\text{select}}$\;

\end{algorithm}

\begin{algorithm}[t]
\caption{Functions for Memory Buffer Update}
\label{alg:functions}
\SetAlgoLined
\SetKwFunction{SelectNewSamples}{SelectNewSamples}
\SetKwFunction{DropOldSamples}{DropOldSamples}
\SetKwFunction{PseudoLabeling}{PseudoLabeling}
\SetKwFunction{ManualLabeling}{ManualLabeling}
\SetKwFunction{UpdateMemoryBuffer}{UpdateMemoryBuffer}
\SetKwFunction{Random}{Random}
\SetKwFunction{TopK}{TopK}
\SetKwFunction{LowK}{LowK}
\SetKwFunction{Num}{Num}
\SetKwProg{Fn}{Function}{:}{end}

\Fn{\SelectNewSamples{$D_t$, $k$, $m^n$}}{
    $X_{t}^{\text{label}} = \{ x_i^n \mid m_i^n \geq 0.5 \}$\;
    \If{\Num($X_{t}^{\text{label}}$) $< k$}{
        $X_{t}^{\text{label}} = X_{t}^{\text{label}} \cup \Random(D_t, k - \Num(X_{t}^{\text{label}}))$\;
    }
    \Else{
        $X_{t}^{\text{label}} = \TopK(D_t, k)$\;
    }
    return $X_{t}^{\text{label}}$\;
}

\Fn{\DropOldSamples{$M_{t-1}$, $k$, $m^o$}}{
    $M_{t-1}^{\text{drop}} = \{ x_i^o \mid m_i^o < 0.5 \}$\;
    \If{\Num($M_{t-1}^{\text{drop}}$) $< k$}{
        $M_{t-1}^{\text{drop}} = M_{t-1}^{\text{drop}} \cup \LowK(M_{t-1}, k - \Num(M_{t-1}^{\text{drop}}))$\;
    }
    \Else{
        $M_{t-1}^{\text{drop}} = \Random(M_{t-1}, k)$\;
    }
    return $M_{t-1}^{\text{drop}}$\;
}

\Fn{\PseudoLabeling{$X_t$, $f(\cdot)$, $k$}}{
    $X'_{t} = $ \SelectNewSamples($D_t$, $k$, $m^n$)\;
    $D'_{t} =$ \ManualLabeling($X'_{t}$)\;
    $X''_t = (X_{t} \setminus X'_{t})$\;
    \For{each $x_i$ in $X''_{t}$}{
        $\hat{y}_i = f(x_i)$ \tcp*[f]{\textcolor{gray}{\textnormal{use detection model to generate pseudo-labels;}}}\ 
    }
    $D''_{t} = \{(x_i, \hat{y}_i) \mid x_i \in X''_{t}\}$\;
    return $D_t = \{D'_t, D''_t\}$\;
}

\end{algorithm}

\subsection{Memory Buffer Update with Strategic Forgetting}

The memory buffer update process is essential for maintaining the relevance and effectiveness of the model in the face of changing distributions. The update strategy depends on the outcome of the drift detection $h(p)$. The detailed process is outlined in Algorithm \ref{alg:memory_update}, and the specific functions used in this process are described in Algorithm \ref{alg:functions}.

According to the result of the sample selection, new samples can be classified into two types: representative ($\mathbf{m}^n \geq 0.5$) or non-representative ($\mathbf{m}^n < 0.5$) for distribution drift. Similarly, old samples can be classified as representative ($\mathbf{m}^o \geq 0.5$) or non-representative ($\mathbf{m}^o < 0.5$) for the common distribution. For non-representative samples, their representative score ($m^n_i$ or $m^o_i$) is not important and all non-representative samples are treated equally. This means that selection among non-representative samples is done randomly.

When no drift is detected, i.e., $h(p) = 0$, the memory buffer is updated based on the labeling resource $k$, which refers to the number of new samples that can get ground truth labeled in each update round, for example, through manual labeling. Essentially, this step involves incorporating $k$ labeled new samples into the memory buffer and discarding $k$ old samples, given the fixed memory size.

To ensure that these labeling resources are used effectively, the process first targets new samples that are representative of the distribution drift, i.e., those with $\mathbf{m}^n \geq 0.5$. Among these representative samples, the top $k$ samples with the highest scores $m^n_i$ are selected. If the number of these representative new samples is less than $k$, all such samples are labeled, and additional samples are randomly selected from the remaining non-representative new samples to fully utilize the labeling resource $k$ (Algorithm \ref{alg:memory_update}, line 2).

To incorporate these selected new samples with ground truth labels into the memory buffer (Algorithm \ref{alg:memory_update}, line 3), an equivalent number of $k$ old samples need to be dropped. 
Old samples that are not representative of the common distribution, identified by \( \mathbf{m}^o < 0.5 \), are prioritized for removal. If the number of such non-representative old samples exceeds \( k \), we randomly select \( k \) samples for removal. If the number is less than \( k \), we drop all non-representative samples and then select additional old samples with the lowest \( m^o_i \) scores to make room for new samples, ensuring that exactly \( k \) samples are removed (Algorithm \ref{alg:memory_update}, line 4).

When drift is detected, i.e. $h(p) = 1$, the strategy involves a more significant update to the memory buffer through strategic forgetting. In the face of significant drift, retaining outdated knowledge is not advisable as it may hinder the model's effective adaptation. Therefore, all old samples that are not representative of the common distribution, i.e., those with $\mathbf{m}^o < 0.5$, are discarded (Algorithm \ref{alg:memory_update}, lines 7-8). The number of new samples added to the buffer is equal to the number of old samples dropped. These new samples are selected using the same method as in the no drift scenario (Algorithm \ref{alg:memory_update}, line 9). 
If the number of non-representative old samples is less than or equal to $k$, all selected new samples can obtain their ground truth labels, similar to the no drift scenario (Algorithm \ref{alg:memory_update}, line 11). 
If the number of old samples dropped exceeds $k$, after selecting and labeling $k$ new samples (Algorithm \ref{alg:functions}, lines 22-23), the remaining new samples, up to the number of old samples dropped, are labeled using pseudo-labels generated by the detection model $f(\cdot)$ (Algorithm \ref{alg:functions}, lines 25-25) to facilitate the continual learning process (Algorithm \ref{alg:memory_update}, line 14). Strategic forgetting, in this context, is advantageous because it allows for the inclusion of a greater number of new samples that naturally represent the updated distribution, which is essential for maintaining the model's accuracy and relevance over time. 

This approach ensures that the buffer is refreshed with the most relevant and representative new samples that accurately reflect the current distribution while discarding outdated non-representative old samples, allowing the model to adapt swiftly to concept drift in a continual learning scenario.

\subsection{Model Fine-Tuning}

This step involves updating/fine-tuning the model using the updated memory buffer/dataset, with a strategy of assigning greater weights to new samples for better adaptation to concept drift. 
Based on the drift detection results $h(p)$, different strategies are employed for each scenario.

For intrusion detection using deep learning-based methods, model training typically involves a task-specific loss \( \mathcal{L}_{\text{task}} \).
This loss can take various forms, such as cross-entropy loss for classification-based approaches \cite{kunang2018automatic,zhao2022tri,zhou2021application} or contrastive loss \cite{aoc-ids,feco} for similarity-based approaches, depending on the specific method used to tackle the problem. 

The task-specific loss \( \mathcal{L}_{\text{task}} \) is generally computed as the sum of individual losses for each sample. 
In a static environment, treating each sample equally is justified since there is no major difference between them.
However, in our dynamic scenario where concept drift occurs, it becomes essential to pay more attention to new samples, as the drift is likely driven by changes in these new samples. Although these new samples may be few, they indeed convey more meaningful information for the model to learn from than the old samples.

In our approach, the memory buffer/dataset after update \( M_t \) consists of both old and new samples, i.e., \( M_t = \{D^o, D^n\} = \{\{x_i^o, y_i^o\}, \{x_i^n, y_i^n\}\} \). We handle the losses of new and old samples differently by assigning a weight to the losses of new samples while keeping the weight for old samples as 1. This can be formulated as:
\[
\mathcal{L}_{\text{task}} = \sum_{(x_i^o, y_i^o) \in D^o} \ell(x_i^o, y_i^o) + \gamma \sum_{(x_j^n, y_j^n) \in D^n} \ell(x_j^n, y_j^n) ,
\]
where \(\ell(x_k, y_k)\) represents the loss for each sample, and \(\gamma\) is a hyperparameter for the weight of the loss for new samples.

This weighting strategy allows us to control the influence of new samples during model training. 
By increasing \(\gamma\) (i.e., setting \(\gamma > 1\)), the model places more emphasis on incorporating these new data points. 
This is particularly beneficial when new samples are a minority in the memory buffer, ensuring they are given adequate importance despite their smaller number. 
This promotes a stronger learning process that adapts to the changing data landscape.

When no drift is detected, i.e. $h(p) = 0$, focusing on preventing catastrophic forgetting is beneficial for maintaining the model's performance. 
To achieve this, we leverage existing regularization-based continual learning methods \cite{lwf}, which introduce constraints on weight updates during fine-tuning to consolidate existing knowledge and mitigate forgetting. 
The overall training loss function, which incorporates both the task-specific and regularization components, is given by:
\[ \mathcal{L}_{\text{no\_drift}} = \mathcal{L}_{\text{task}} + \lambda \cdot \mathcal{L}_{\text{reg}}, \]
where \( \mathcal{L}_{\text{reg}} \) is the regularization loss that helps preserve important weights, and \( \lambda \) is a hyperparameter that balances the contributions of the task-specific loss and the regularization term. This approach ensures that while the model learns new tasks, it does not forget how to perform previously learned tasks, thereby maintaining its overall effectiveness when there is no significant drift. 

When drift is detected, i.e. \( h(p) = 1 \), the model must prioritize rapid adaptation to the new distribution over the consolidation of previous knowledge. 
By excluding the regularization term, the model can fully focus on learning the new distribution, which is crucial for adapting quickly to the detected drift. The loss function, in this case, simplifies to:
\[ \mathcal{L}_{\text{drift}} = \mathcal{L}_{\text{task}} \]
This streamlined approach ensures that the model's training prioritizes the adaptation to the new distribution, thereby enhancing its ability to respond to significant drift effectively.

These four steps, drift detection, sample selection, memory buffer update, and model fine-tuning, ensure swift and accurate adaptation to concept drifts, sustaining model performance despite the non-stationary nature of real-world data.

\section{Experiments}
\label{sec:experiments}

This section presents the experimental setup and results. We detail the datasets, experiment settings, and baseline methods in \subsectionautorefname{~\ref{sec:dataset}}. The superior performance of SSF over the baseline methods in the online setting is demonstrated in \subsectionautorefname{~\ref{sec:comparative experiments}}. Additionally, ablation experiments are conducted to illustrate the contribution of sub-components in SSF in \subsectionautorefname{~\ref{sec:ablation experiments}}. 
The experimental results reveal that our method outperforms the state-of-the-art (SOTA) techniques and underscores the significant contributions of each component in our system to its capacity and adaptability.

\subsection{Dataset Preparation and Experiment Settings}
\label{sec:dataset}

\subsubsection{Datasets}

    We conduct experiments on two datasets that are widely used in network intrusion detection \cite{feco,conIDS,pajouh2017two}, i.e., NSL-KDD and UNSW-NB15. 

    The NSL-KDD dataset comprises 125,973 training samples and 22,544 test samples of network traffic. The UNSW-NB15 dataset is divided into a training set with 175,341 samples and a test set with 82,332 samples.

\subsubsection{Baselines}
\label{sec:baselines}

    We compare our proposed method with five well-established continual learning methods within the same continual learning setting. The continual learning baselines include AOC-IDS \cite{aoc-ids}, which is specifically designed for the intrusion detection domain, as well as other notable continual learning approaches including Experience Replay (ER) \cite{er1,er2}, Averaged Gradient Episodic Memory (AGEM) \cite{agem}, Learning without Forgetting (LwF) \cite{lwf}, and Elastic Weight Consolidation (EWC) \cite{ewc}. 
    
    \begin{itemize}
        \item \textbf{AOC-IDS} enhances the memory buffer with pseudo-labeled samples from the incoming stream, and introduces label perturbation to prevent overfitting. 
        \item \textbf{ER} uses reservoir sampling to randomly select samples from the incoming data stream, which are then incorporated into the memory buffer.
        \item \textbf{AGEM} stores past gradients and projects new task gradients to minimize interference between tasks. 
        \item \textbf{LwF} penalizes the model for deviating from the outputs it produced on earlier tasks by knowledge distillation.
        \item \textbf{EWC} estimates the importance of each parameter using the Fisher Information Matrix and constrains the update of important parameters from previous tasks.
    \end{itemize}

    We also include a static model without continual learning for reference to quantify the improvements in model performance due to continual learning in the context of concept drift.

    \subsubsection{Experiment Settings} 
    In a continual learning scenario, the IDS starts with a limited number of labeled datasets. The rest of the samples are incrementally fed into the system without ground truth labels in a streamlined manner. 
    Following the widely used setting, we select 20\% of the data from the original training dataset as the initial labeled dataset, which amounts to 25,194 samples for NSL-KDD and 38,068 samples for UNSW-NB15, for initial training and memory buffer initialization. 
    The performance of all methods is evaluated on the test dataset where concept drift occurs. 
    All comparative continual learning methods utilizing memory buffers adhere to this fixed memory buffer size for storing data. 
    The IDS undergoes updates after every incremental addition of 5,000 samples for NSL-KDD and 20,000 samples for UNSW-NB15.

    To validate the method’s universality across different models, we use different models for the NSL-KDD and UNSW-NB15 datasets. 
    For NSL-KDD, we use the detection module backbone in the AOC-IDS \cite{aoc-ids}, which consists of an Autoencoder (AE) for representation learning, followed by a Gaussian fit process for labeling the input. 
    The AE architecture for NSL-KDD is [121, 64, 32, 64, 121]. 
    For UNSW-NB15, we employ a simple fully connected neural network with the architecture [196, 128, 64, 128, 196, 1], where the final layer acts as a classifier to output the labels. 
    The models are trained using a Stochastic Gradient Descent optimizer with a learning rate of 0.001 and a batch size of 128. 
    The training process is optimized using the most popular contrastive loss, InfoNCE \cite{InfoNCE}. 
    When no drift is detected, we incorporate regularization-based continual learning, specifically LwF, in our approach. 
    To ensure fairness, we applied the same detection models and training processes uniformly across all comparative methods.
    The reported results are the average of five consecutive rounds of experimentation, guaranteeing their reliability.

\begin{table}[!t]
\caption{Performance (\%) comparison between SSF and baselines. The highest metric performance is bolded.}
\resizebox{\columnwidth}{!}{%
\begin{tabular}{ccccccccc}
\toprule
\multirow{2}{*}{Method} & \multicolumn{4}{c}{NSL-KDD} & \multicolumn{4}{c}{UNSW-NB15} \\ \cmidrule(lr){2-5} \cmidrule(lr){6-9}
                        & Acc.   & Pre.  & Rec.  & F1  & Acc.    & Pre.   & Rec.   & F1   \\ \midrule
\textbf{SSF (Ours)}    & \textbf{90.50}  &89.22  & \textbf{94.79} &   \textbf{91.90} &\textbf{90.27}  & 89.06   & 93.89    &\textbf{91.40}      \\
AOC-IDS     &   81.74    &  78.97     &  92.55  & 85.21 & 81.50   &   75.07     &  \textbf{99.41}     &  85.54     \\
ER           &   86.52     &  91.15     &   84.60    &  87.65   &   83.66   &   77.88    & 98.23      &   86.88      \\
AGEM        &  84.21 &  \textbf{92.86}  &  78.40  &  84.81 &  84.86 &  79.15  &  98.44  &  87.74    \\
LwF               &  82.65  &  89.23  &  79.06  &  83.78 &  86.43 &  88.25  &   87.02  &  87.59  \\
EWC               &  81.70  &  89.14  &  77.38  &  82.71 &  86.55  &  \textbf{89.56}  &  85.73  &  87.52    \\
Static            &   84.48 &  88.94  &  83.45  &  86.06 &    81.75    & 75.42 &   99.17    &  85.68    \\ \bottomrule
\end{tabular}}
\label{tab:compare}
\vspace{-0.5cm}
\end{table}

\subsection{Comparitive Experiments}
\label{sec:comparative experiments}
    We compare SSF against five well-established continual learning methods, as well as a static model for reference. All methods are evaluated under the same continual learning setting, where we can only obtain 1\% ground truth labels of new unlabeled samples. 
    Moreover, we highlight the exceptional capacity of SSF in scenarios with extremely limited labeling resources by analyzing its performance across different labeling conditions.
    Additionally, we provide a visualization example to illustrate how SSF enhances the model's ability.
    
    \subsubsection{Overall Performance}
        As displayed in \tablename{~\ref{tab:compare}}, SSF surpasses other methods in both accuracy and F1 score on the NSL-KDD and UNSW-NB15 datasets. Specifically, our method achieves high detection accuracies of 90.50\% and 90.27\% on the NSL-KDD and UNSW-NB15 datasets, respectively, demonstrating improvements of approximately 4.0\% and 3.7\%. 
        Notably, our method is the only one to perform uniformly well on both datasets, while other compared methods either exhibit generally poor performance across both datasets or excel only on one dataset. 

        AOC-IDS is designed under the assumption of infinite memory capacity, allowing it to retain all past samples while incorporating new ones. 
        This overlooks the inherent constraints of physical storage, which have finite capacities and cannot store unlimited amounts of data. In a realistic continual learning scenario with finite memory constraints, AOC-IDS loses its effectiveness, indicating that AOC-IDS relies heavily on having access to the entire historical dataset. It lacks the ability to select representative samples and learn rich, concept drift-adaptive information from a limited set of samples. 
        
        Both ER and AGEM are memory-based continual learning methods that store and replay a subset of past information from previous tasks in a fixed-size memory buffer, similar to our approach. 
        The performance of ER shows improvement compared to the static, non-updated scenario, successfully achieving learning new knowledge without catastrophically forgetting old knowledge. However, this conservative and simple random selection approach is highly likely to miss the most representative samples and fail to capture the most critical samples for adapting to concept drift, preventing it from achieving exceptional performance. 
        Similarly, the moderate performance of AGEM demonstrates that AGEM's approach of preserving past knowledge by gradient projection is far less effective than our method’s approach of selecting the most representative samples for drift adaptation. Our method prioritizes learning new, useful knowledge over merely preserving old knowledge, which proves to be more beneficial. 

\begin{figure*}[!t]
    \centering
    \begin{subfigure}{0.46\textwidth}
        \centering
        \includegraphics[width=\textwidth]{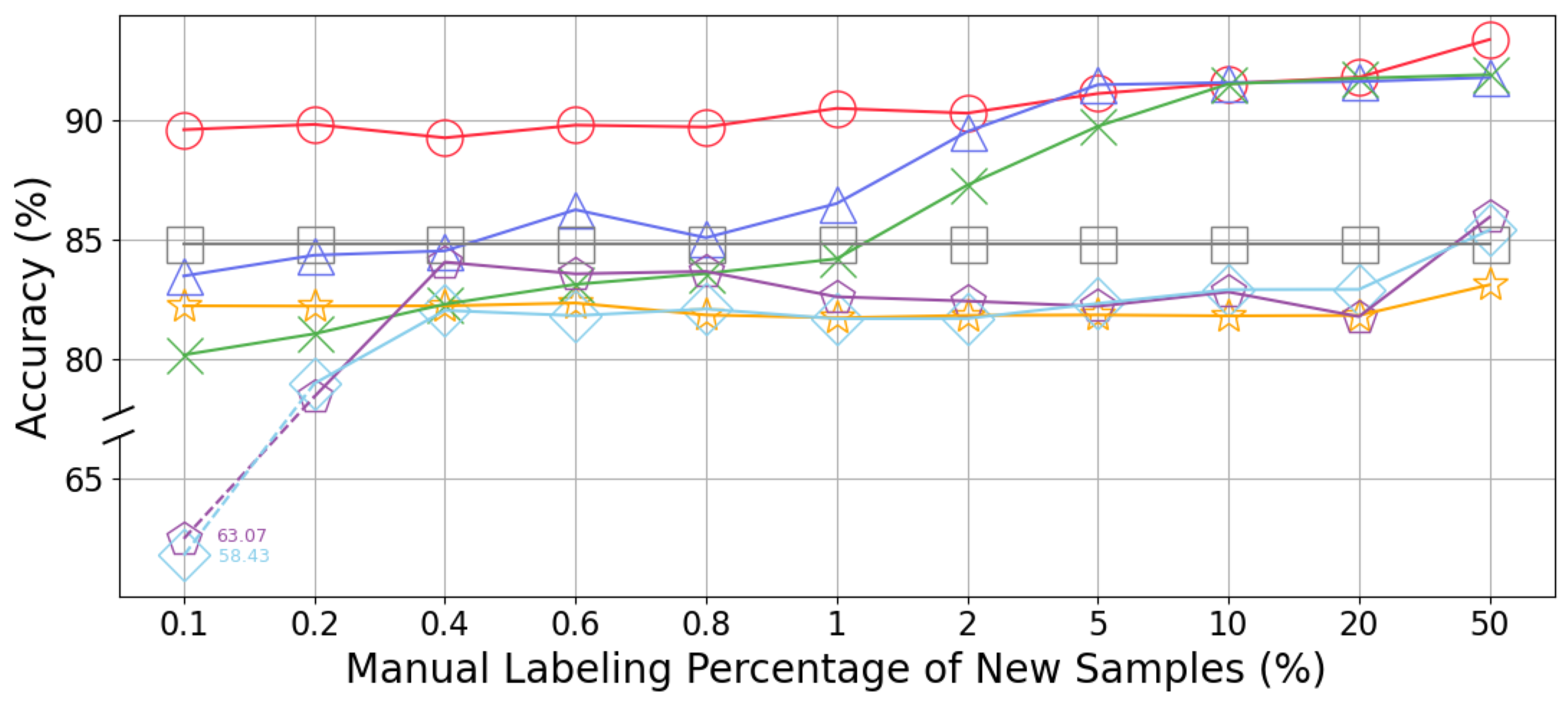}
        \caption{NSL-KDD Dataset}
        \label{fig:nsl_label}
    \end{subfigure}%
    \begin{subfigure}{0.46\textwidth}
        \centering
        \includegraphics[width=\textwidth]{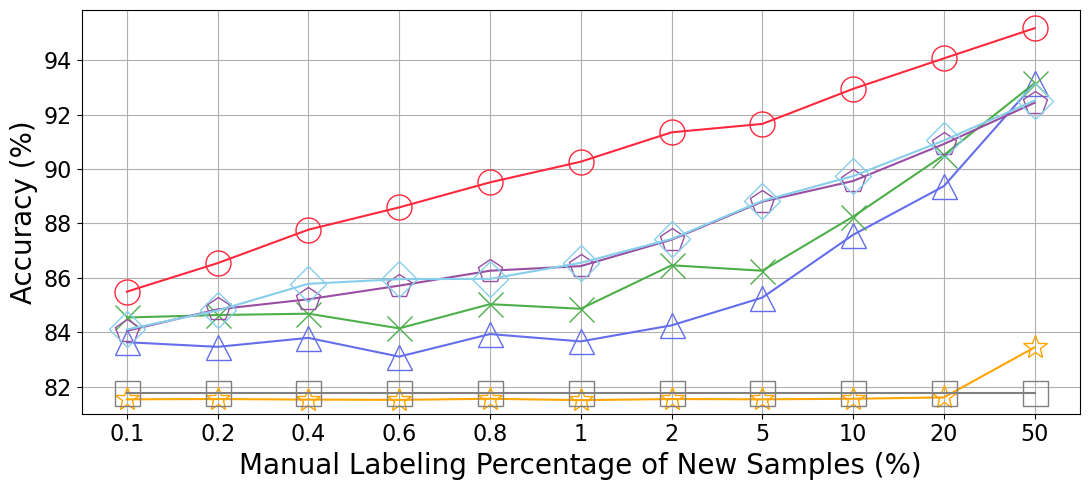}
        \caption{UNSW-NB15 Dataset}
        \label{fig:unsw_label}
    \end{subfigure}
    \begin{subfigure}{0.074\textwidth}
        \centering
        \includegraphics[width=\textwidth]{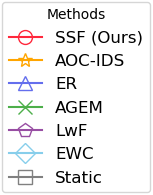}
        \label{fig:legend}
        \vspace{1.85cm}
    \end{subfigure}
    \caption{Performance (accuracy) comparison under different labeling resources.}
    \label{fig:comparison}
    \vspace{-0.5cm}
\end{figure*}

        LwF and EWC, which are regularization-based methods, perform relatively well on the UNSW-NB15 dataset but poorly on the NSL-KDD dataset. These methods do not require retaining past data; instead, they use constraints to mitigate catastrophic forgetting. However, they face significant limitations when the absolute number of new samples available is very small.
        To be specific, in the NSL-KDD dataset, the model is updated every 5000 samples, providing only 50 new samples (1\%) per update. In contrast, the UNSW-NB15 dataset allows for model updates every 20,000 samples, providing 200 new samples (1\%) per update. The poor performance of these methods on NSL-KDD can be attributed to their inability to effectively learn from such a small number of new samples, which hampers their ability to update knowledge without disrupting existing information.
        In contrast, by focusing on identifying and utilizing the most representative samples, we maximize the value of labeled resources. With the help of old samples representing the common distribution for new patterns, our method proves to be particularly advantageous in scenarios with very limited labeling resources.

    \subsubsection{Accuracy under Different Label Conditions}
    
        We analyze the accuracy of our method in comparison to other baseline methods under varying label resource conditions, as illustrated in \figurename{~\ref{fig:comparison}}. It is evident that continual learning is most effective when only a small portion of new samples are labeled. If the majority of new samples are manually labeled, the IDS essentially loses its purpose as an automated detection method. The continual learning method that can operate effectively with fewer labeled resources is superior and more practical. 
        Consequently, Fig. 4 primarily focuses on scenarios with less than 1\% labeled resources. Higher labeling percentages, up to 50\%, are also provided for reference.

        For the NSL-KDD dataset, our method performs exceptionally well across all labeling resource conditions, achieving nearly 90\% accuracy even with just 0.1\% labeled data. We consistently outperform other baselines when labeling resources are under 2\%. Beyond this level of labeling resources, other memory-based methods like ER and AGEM achieve comparable performance. However, in terms of the labeling resources needed to achieve the same performance level, our method reaches the performance level that others achieve with 5\% labeling resources, using only 0.1\% of the resources. This underscores a nearly 50-fold reduction in labeling resource requirements of our approach. 
        For the UNSW-NB15 dataset, our method consistently outperforms other baselines across all labeling percentages. Our method reaches the performance level that others achieve with 20\% labeling resources, using only 1\% of the resources. This demonstrates a nearly 20-fold reduction in labeling resource requirements.
        
        Overall, our method demonstrates superior performance across both datasets and labeling resource conditions. It highlights the effectiveness of our techniques to optimize the use of the memory buffer and maximize the value of limited labeling resources by distinguishing and storing the most representative samples to improve the model's adaptability.

\begin{figure}[!t]
    \centering
    
    \begin{subfigure}{0.25\textwidth}
        \centering
        \includegraphics[width=\textwidth]{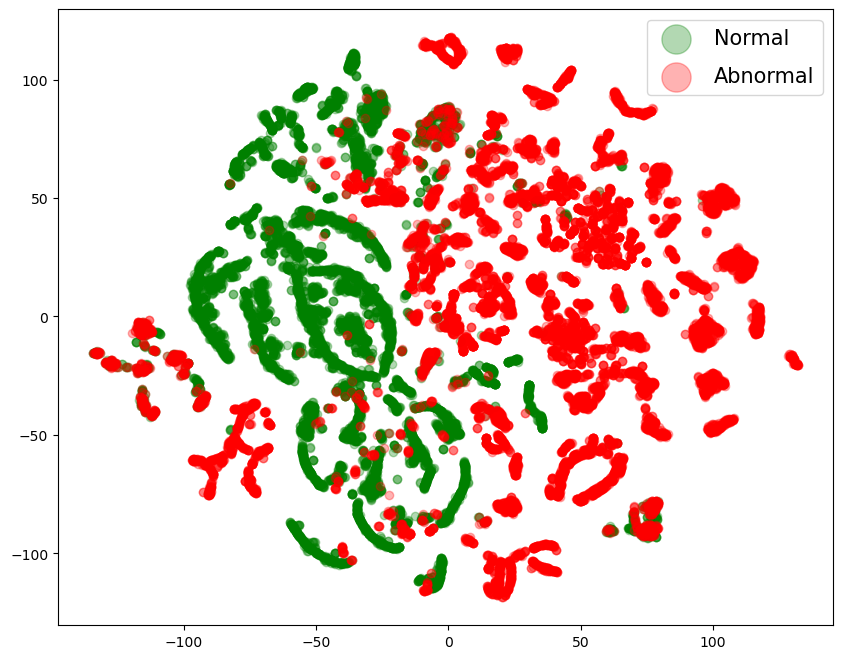}
        \caption{SSF (Ours)}
        \label{fig:our_vis}
    \end{subfigure}%
    \begin{subfigure}{0.25\textwidth}
        \centering
        \includegraphics[width=\textwidth]{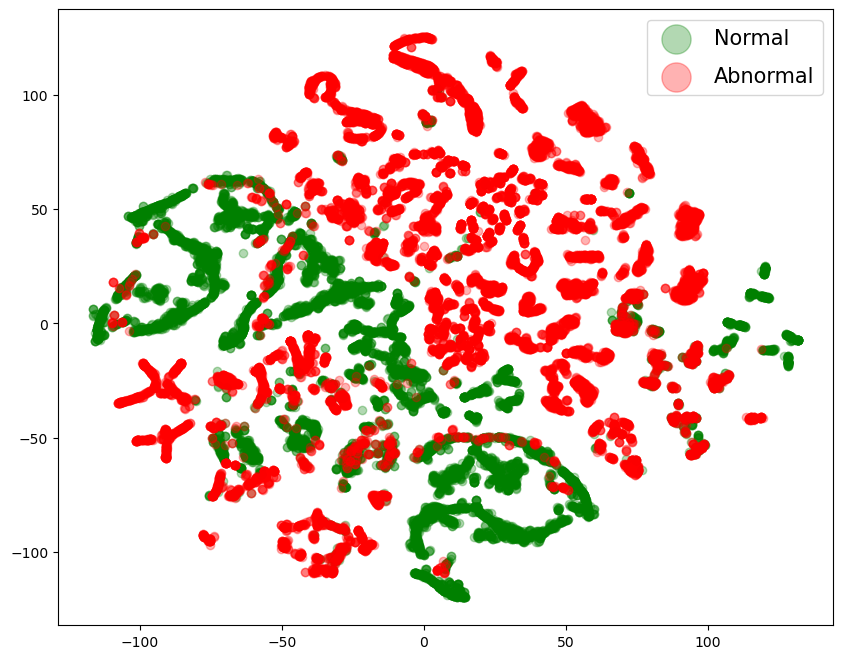}
        \caption{AOC-IDS}
        \label{fig:aoc_vis}
    \end{subfigure}
    \caption{t-SNE visualization of decoder feature extraction on NSL-KDD dataset.}
    \label{fig:visualization}
    \vspace{-0.5cm}
\end{figure}

    \subsubsection{Visualization}
        \figurename{~\ref{fig:visualization}} shows the t-SNE visualization of the features extracted by the decoder on the NSL-KDD dataset. The visualization illustrates the distribution and clustering of the features after dimensionality reduction. For IDS, the goal is to clearly distinguish between normal samples and abnormal samples. Therefore, an optimal model will show clear boundaries and minimal overlap between the clusters of normal and abnormal samples.
        In \figurename{~\ref{fig:our_vis}}, the visualization for AOC-IDS shows significant overlap between normal (green) and abnormal (red) samples. The normal regions are largely covered by abnormal data, reflecting its struggle to maintain detection ability under its own continual learning update framework.
        In contrast, \figurename{~\ref{fig:aoc_vis}} illustrates that our method achieves a more distinct separation between the clusters of normal and abnormal samples. The clear separation of the clusters suggests that the model has learned more useful features for distinguishing between the two classes.

\subsection{Ablation Experiments}
\label{sec:ablation experiments}

    In this section, we perform ablation experiments to evaluate the effectiveness of key techniques of SSF, including strategic sample selection and strategic forgetting. 
    Additionally, we conduct an ablation study with and without (w/o) using the existing regularization-based continual learning method, specifically LwF, to illustrate that LwF is not the primary reason for the improved performance of our method. 

    \subsubsection{Strategic Sample Selection} We compare our strategic sample selection algorithm with simple random sampling, denoted by `w/o sample selection'. As shown in \tablename{~\ref{tab:ablation}}, the results demonstrate that our strategic sample selection algorithm significantly outperforms random sampling. 
    This indicates that merely selecting samples randomly is likely to miss crucial information, especially when the proportion of new samples available for labeling is very low. Randomly selected samples may not adequately represent the new distribution. 
    In contrast, by using our strategic sample selection algorithm, even with a limited number of new samples, the most representative samples can still effectively represent the new distribution, particularly the drifted part. 
    By maximizing the value of labeled resources, our approach leads to notable performance enhancements compared to random sampling.

    \subsubsection{Strategic Forgetting}
    We compare adopting strategic forgetting, which discards all non-representative old samples when drift is detected, with dropping the same number of samples as labeling resources the whole time, described as `w/o strategic forgetting'. 
    \tablename{~\ref{tab:ablation}} suggests that the adoption of strategic forgetting enhances performance. This improvement is due to the timely removal of outdated or potentially incorrect knowledge, ensuring the model adapts effectively to new patterns and frees up space in the limited memory buffer for incorporating more recent data. During concept drift, retaining new samples that naturally represent the new pattern can be more beneficial for adaptation than keeping old samples, thereby facilitating adaptation.

    \subsubsection{Regularization}
    We compare adopting LwF when no drift is detected, with no adoption of such regularization-based continual learning method, described as `w/o regularization'. 
    \tablename{~\ref{tab:ablation}} demonstrates that the adoption of LwF enhances performance, but the improvement is minimal. Therefore, this proves that LwF is not the primary reason for the improved performance.
    Instead, it is the strategic sample selection and strategic forgetting that predominantly account for the significant improvements observed in our system's performance.

\begin{table}[!t]
\caption{Ablation study results for SSF. The highest metric performance is bolded.}
\resizebox{\columnwidth}{!}{
\begin{tabular}{ccccccccc}
\toprule
\multirow{2}{*}{Method} & \multicolumn{4}{c}{NSL-KDD} & \multicolumn{4}{c}{UNSW-NB15} \\ \cmidrule(lr){2-5} \cmidrule(lr){6-9}

& Acc.   & Pre.  & Rec.  & F1  & Acc.    & Pre.   & Rec.   & F1   \\ \midrule
SSF (Ours)  & \textbf{90.50}  &89.22  & \textbf{94.79} &   \textbf{91.90} & \textbf{90.27}  & \textbf{89.06}   & 93.89    &\textbf{91.40}    \\
w/o sample selection     &   87.83    &   \textbf{91.33}   &    86.92   &  88.99  &   86.57     &   82.29    &  \textbf{96.39}     &   88.77  \\
w/o strategic forgetting    &  89.09 &   90.50   &    90.34  &  90.38   &    89.32  & 86.57   & 95.51    & 90.79   \\
w/o regularization         & 90.20       & 89.04    &  94.43    & 91.64    & 90.00      & 88.04      & 94.79     & 91.21    \\
\bottomrule
\end{tabular}}
\label{tab:ablation}
\vspace{-0.5cm}
\end{table}

\section{Related Work}
In this section, we first discuss the existing continual learning methods. Then, we introduce continual learning targeted for intrusion detection to explain the research gap.

\subsection{Continual Learning}
Continual learning strategies are generally categorized into memory-based \cite{rolnick2019experience }, regularization-based \cite{ahn2019uncertainty}, and expansion-based approaches \cite{ye2023self}. We focus on the first two categories since expansion-based methods, which involve expanding the model architecture, do not align with our scenario with a fixed model structure. 
Therefore, we focus on comparing our method with memory-based and regularization-based approaches to ensure a fair evaluation.

\subsubsection{Memory-based Methods}
Memory-based methods address concept drift and mitigate catastrophic forgetting by storing and replaying a subset of past information from previous tasks, including but not limited to data, such as ER \cite{er1,er2} and gradients, such as GEM (Gradient Episodic Memory) \cite{gem} and A-GEM \cite{agem}, an improved version of GEM that simplifies the computation by using average gradients. 
However, existing methods overlook the critical `drift' aspect of concept drift and fail to recognize the value of strategic forgetting the potential benefits of strategic forgetting, missing the potential for it to facilitate rapid adaptation in dynamic environments.

\subsubsection{Regularization-based Methods}
Regularization-based methods \cite{lwf,ewc} preserve knowledge from prior tasks while learning new tasks by applying constraints to the learning process. 
These methods typically incorporate a regularization term into the objective function, penalizing updates that would change or overwrite weights crucial for past tasks. 
Although these methods do not require storing previous data, they often struggle when labeling resources are very limited due to insufficient learning from available new data samples and an over-reliance on the initial training.

\subsection{Continual Learning for Intrusion Detection}

While continual learning mechanisms have been adopted into IDSs to accommodate dynamic environments, significant issues remain unaddressed. 
For example, AOC-IDS \cite{aoc-ids} periodically updates the IDS to address concept drift but retains all encountered data, neglecting practical memory buffer limitations. 
Similarly, the approach in \cite{han2021log} necessitates labor-intensive labeling for continuous training, overlooking the constraints on labeling resources. 
Han et al. \cite{owad} proposed a continual learning mechanism that targets labeling for the most influential samples, thereby reducing the labeling burden with a fixed memory buffer size. However, this method is tailored for zero-positive learning, focusing solely on training with normal data, which leads to a waste of valuable abnormal data and labeling resources since new abnormal samples are discarded after manual labeling.

In this study, we introduce a novel memory-based continual learning approach that optimally leverages both normal and anomalous samples within realistic constraints such as fixed memory size and limited labeling resources, thereby addressing the aforementioned challenges.

\section{Conclusion}
In this paper, we introduced a novel 
continual learning method for IDSs, comprising four steps: drift detection, sample selection, memory buffer update, and model fine-tuning. 
The drift detection step lays the foundation for the following steps, which handle drift and no-drift scenarios differently. The sample selection step identifies the most representative new samples while achieving minimal labeling cost. Strategic forgetting, used in the memory buffer update step, ensures the model aligns with the current data. By assigning weights to losses of new samples during fine-tuning, we balance learning from new tasks and consolidating previous knowledge.
These four steps ensure that IDSs can swiftly and accurately adapt to concept drifts with minimal cost. 
Experimental results on the NSL-KDD and UNSW-NB15 datasets demonstrate the superior performance and adaptability of our method, surpassing state-of-the-art solutions. The optimal labeling efficiency of our approach is further highlighted under extremely limited labeling resource conditions. Furthermore, our ablation study confirms the contributions of our proposed techniques. Overall, our proposed method makes IDSs more robust and responsive in dynamic network environments, balancing efficient labeling with rapid adaptation to concept drifts.

\section*{Acknowledgment}
This research is supported in part by the National Natural Science Foundation of China (Grant No. 92067109, 61873119, 62211530106), in part by the Shenzhen Science and Technology Program (Grant No. ZDSYS20210623092007023, GJHZ20210705141808024), and in part by the UGC General Research Fund (Grant No. 17203320, 17209822) from Hong Kong.

\bibliographystyle{IEEEtran}
\bibliography{references}

\end{document}